\documentclass[letterpaper, 10 pt, conference]{IEEEtran}  

\IEEEoverridecommandlockouts                              

\usepackage{hyperref}
\usepackage{multirow}
\usepackage{geometry}
\usepackage{graphicx}
\usepackage{subcaption}
\usepackage{placeins}
\usepackage[nolist,nohyperlinks]{acronym}
\usepackage{cite}
\usepackage[none]{hyphenat}
\title{\vspace{6mm}An Affective Robot Companion for Assisting \\ the Elderly in a Cognitive Game Scenario}

\author{
\IEEEauthorblockN{Nikhil Churamani, Alexander Sutherland and Pablo Barros} 
\IEEEauthorblockA{
Knowledge Technology, Department of Informatics\\
University of Hamburg, Hamburg, Germany\\
Email:\{5churama, sutherland, barros\}@informatik.uni-hamburg.de}
}

\begin{document}

\maketitle
\thispagestyle{empty}
\pagestyle{empty}

\begin{abstract}
Being able to recognize emotions in human users is considered a highly desirable trait in \ac{HRI} scenarios. However, most contemporary approaches rarely attempt to apply recognized emotional features in an active manner to modulate robot decision-making and dialogue for the benefit of the user. In this position paper, we propose a method of incorporating recognized emotions into a \ac{RL} based dialogue management module that adapts its dialogue responses in order to attempt to make cognitive training tasks, like the 2048 Puzzle Game, more enjoyable for the users.

\end{abstract}


\section{Introduction}
\label{sec:introduction}

Decreased physical, mental and social activity may lead to several cognitive disorders in the elderly. As an individual retires from work, the risk of cognitive-decline increases with age due to decreased mental and physical activity~\cite{Rohwedder2010}. Engaging in cognitively stimulating activities such as playing board games or solving puzzles may reduce the risk of severe cognitive deterioration and protect against the development of dementia~\cite{Glei2005}. Furthermore, coping with the decreased social activity with the inclusion of robot companions may lead to positive long-term benefits as a result of modelling engaging human-robot interactions~\cite{leite2013social}. Such robot companions can assist the users in solving cognitive tasks such as puzzles, keeping them mentally stimulated and at the same time engaging them in meaningful affective interactions helping them to cope with the stress of their day-to-day life. The goal of such a robot thus becomes to improve the performance of the user in the task and at the same time reduce their stress, improving the overall quality of their daily life.  

Accounting for the user's affective state becomes a critical part of such an interaction design. Learning to adapt to the user's \textit{mood} allows the robot to improve their interaction experience by changing its strategy, given the user is feeling positive or negative while solving the task. Such an affective understanding in the robot forms the basis for learning different emotional expressions in order to empathise with the user~\cite{Churamani2018Learning}. Additionally, this affective appraisal in the robot can also be used to model an interaction with the user, directed towards solving a specific task. 

In this position paper, we attempt to model a pedagogical approach towards teaching the user to solve a cognitive task that requires logical and spacial awareness. The robot shall actively monitor the emotional state of the user and use it, along with the performance of the user in the task, to adapt its strategy towards modelling an interaction with the user. 

\begin{figure}[t]
    \centering
    \includegraphics[width=0.95\columnwidth]{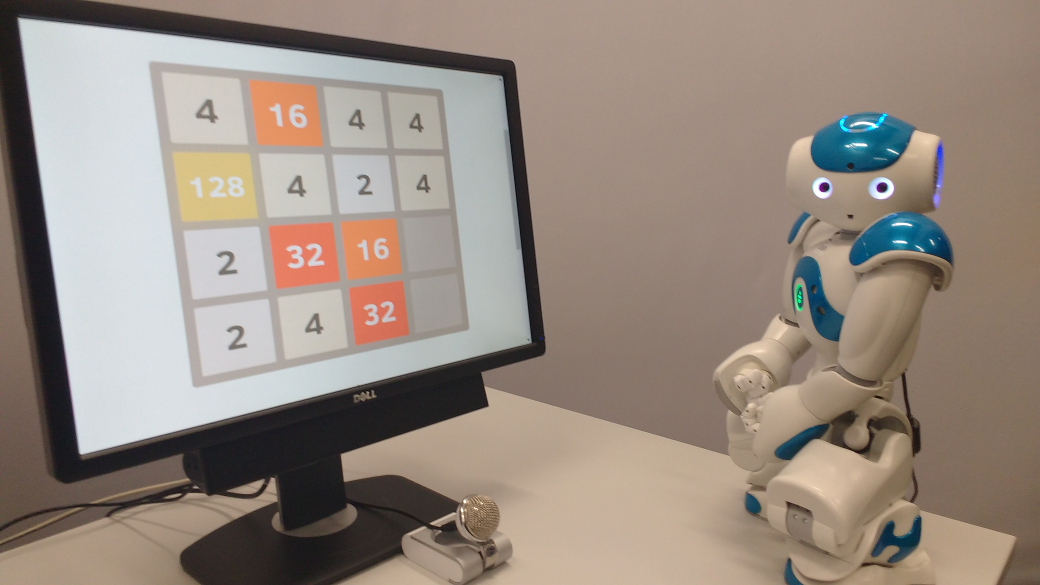}
    \caption{Set-up: The NAO robot as an advisory companion using visual, auditory and linguistic cues to infer the emotional state of the user, offering assistance in the $2048$ Puzzle Game.}
    \label{fig:experiment}
\end{figure}

\begin{figure*}
\centering
\includegraphics[width=0.7\textwidth]{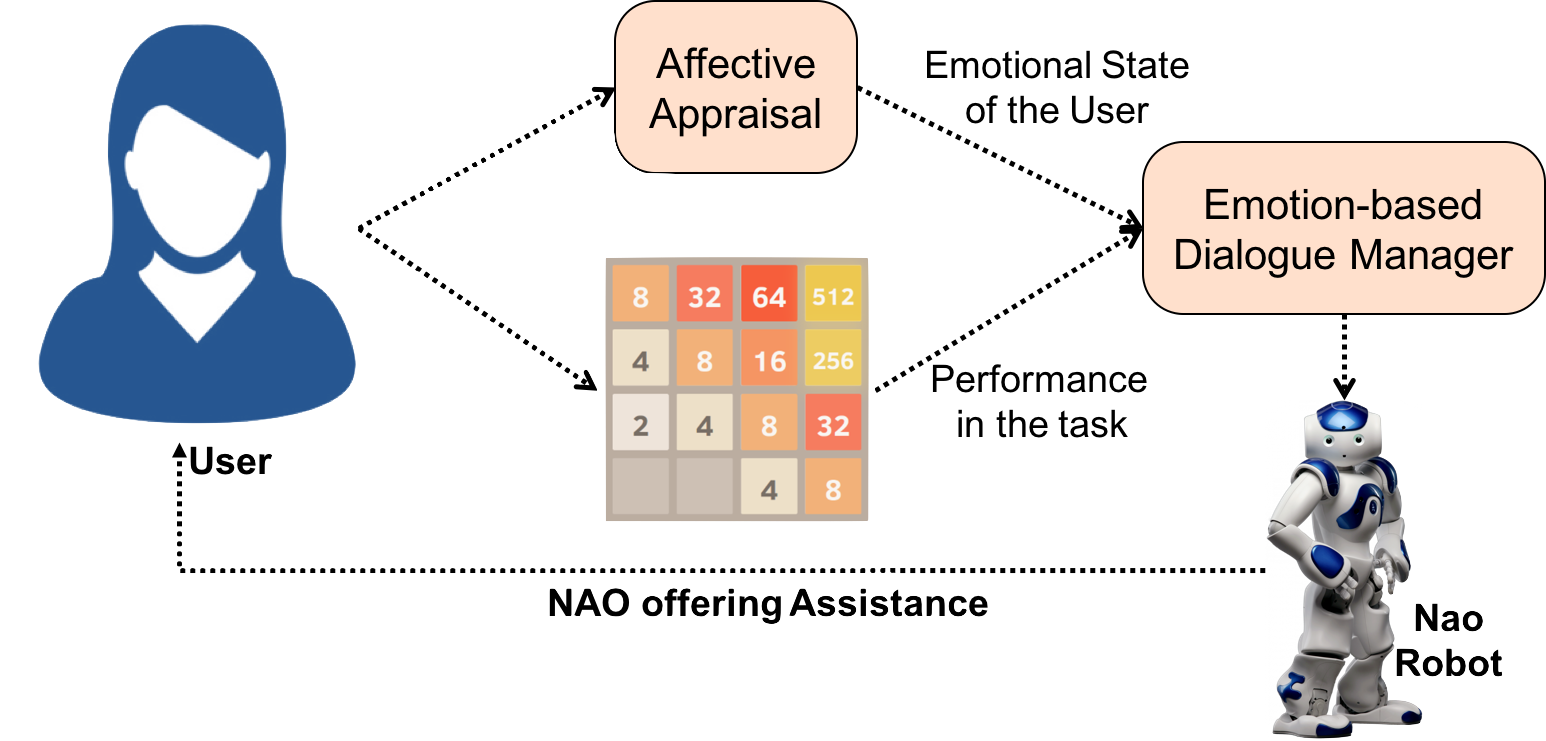}
\caption{User asking the Nao Robot for assistance in solving the 2048 Puzzle Game scenario.}
\label{fig:model}
\end{figure*}

\section{Interaction Scenario}
\label{sec:scenario}

The interaction scenario shall involve the subjects playing several rounds of the $2048$ Puzzle\footnote{https://gabrielecirulli.github.io/2048/ [Accessed 01.04.2018]
} Game~\cite{Neller2015} with the Nao robot (see Fig.~\ref{fig:experiment}). The robot shall be pre-trained (using \acf{RL} techniques~\cite{mnih2015humanlevel}) on solving the game and shall provide suggestions to the users, depending upon their emotional state, for the next best move to succeed in the game. If the user is feeling positive and performing well in the game, the robot shall focus on further improving their performance, actively giving feedback (verbal and gestural) about their moves in the game. In case the user is feeling negative or frustrated, the robot shall focus more on the user and try to improve their emotional state. It shall adapt its behaviour to motivate the user towards the task by offering them assistance and help them cope with the frustration.
The overall goal of the robot, in either situation, shall be to complete the game with the user and at the same time achieve an overall improvement in the mood of the user. The robot should also learn when to give advice to the user, making sure it does not trivialise the task for the user and allows them to make their own decisions. This shall be achieved by monitoring both the performance and the emotional state of the user.

The scenario shall be realised using an emotion-based dialogue management system that shall learn to interact with the user depending upon their emotional state. The emotion perception and intrinsic emotion models shall be based on the works of \emph{Churamani et al.}~\cite{Churamani2018Learning} realising multi-modal sensory capabilities, adding linguistic information to the model, to infer the emotional state of the user. This affective understanding, along with the performance of the user in the task, shall be used to model an emotional dialogue with the user. 

To measure the performance of the model and to study the impact of using the affective understanding of the robot to adapt its ability to assist the user, the model shall be compared to a baseline condition where the robot offers assistance to the users whenever they ask for it. In such a scenario, the robot does not model the emotional state of the user. Any time the users asks for assistance (a hint), the robot provides the same. The proposed scenario can be seen in Fig~\ref{fig:model}.


\section{Expected results}
\label{sec:results}
It is expected that the affective robot companion, that can adapt its ability to model an interaction with the user based on their emotional state and performance in the task, is perceived as more enjoyable and mindful than a static robot that passively waits for user queries. This is expected to lead to users playing the game longer, therefore having a positive impact on their cognitive abilities by training longer. Furthermore, this work intends to show that users who play with the affective robot shall, on average, perform better over time than baseline users as the robot`s mixed initiative approach towards the interaction, intervening and offering assistance in the task, helps users to faster develop a functional game strategy.

Experiment evaluation criteria shall be based on the works of Ho and MacDorman~\cite{ho2010revisiting} and Gray et al.~\cite{gray2007dimensions}, as well as objective performance evaluations based on emotion recognition and the improvement in the performance of the users, over time.



\section{Conclusion}
\label{sec:conclusion}
In this position paper, an emotion-based dialogue management model is envisaged that uses reinforcement learning techniques to learn optimal dialogue states based on the emotional state of the user. The model is proposed to be able to infer the emotional state of the user, based on visual, auditory and linguistic information, and use this affective understanding to assist the users towards solving the $2048$ Puzzle Game. 


\section*{Acknowledgement}

The authors gratefully acknowledge partial support from the German Research Foundation DFG under project CML (TRR 169) and the European Union under project SOCRATES (No. 721619).

\begin{acronym}
\acro{MLP}{Multilayer Perceptron}
\acro{CNN}{Convolutional Neural Network}
\acro{HRI}{Human-Robot Interaction}
\acro{RL}{Reinforcement Learning}
\end{acronym}

\bibliographystyle{IEEEtran}

\bibliography{Pos-IJCNN.bib}

\end{document}